\documentclass[runningheads]{llncs}

\usepackage{float}
\usepackage{graphicx}
\usepackage{longtable} 
\usepackage{comment}
\usepackage[export]{adjustbox}
\usepackage{float}
\usepackage{amssymb}
\usepackage{array}
\usepackage{xcolor} 

\begin{document}




%
\title{Client Error Clustering Approaches in Content Delivery Networks (CDN) } 
%
%
\author{Ermiyas Birihanu\inst{1} \and
Jiyan Mahmud \inst{1} \and
P\'eter Kiss, \inst{1} \and 
Adolf  Kamuzora\inst{1}\and
Wadie Skaf\inst{1} \and 
Tom\'a\v{s} Horváth\inst{1} \and 
Tam\'as Jursonovics, \inst{2} \and 
Peter Pogrzeba, \inst{2} \and 
Imre Lend\'ak\inst{1}
}
\authorrunning{E. Birihanu et al.}
%
\institute{Telekom Innovation Laboratories, Data Science and Engineering Department,
Faculty of Informatics, Eötvös Loránd University
Pázmány Péter str. 1/A, 1117 Budapest, Hungary
\\
\email{\{ermiyasbirihanu, jiyan, axx6v4, adolfnfsp, skaf, tomas.horvath lendak\}@inf.elte.hu }, \\ home page:
\texttt{http://t-labs.elte.hu/}
\and
\ Deutsche Telekom, Berlin, Germany}

%
\maketitle              
\begin{abstract}
Content delivery networks (CDNs) are the backbone of the Internet and are key in delivering high quality video on demand (VoD), web content and file services to billions of users. CDNs usually consist of hierarchically organized content servers positioned as close to the customers as possible. CDN operators face a significant challenge when analyzing billions of web server and proxy logs generated by their systems. The main objective of this study was to analyze the applicability of various clustering methods in CDN error log analysis. We worked with real-life CDN proxy logs, identified key features included in the logs (e.g., content type, HTTP status code, time-of-day, host) and clustered the log lines corresponding to different host types offering live TV, video on demand, file caching and web content. Our experiments were run on a dataset consisting of proxy logs collected over a 7-day period from a single, physical CDN server running multiple types of services (VoD, live TV, file). The dataset consisted of ~2.2 billion log lines. Our analysis showed that CDN error clustering is a viable approach towards identifying recurring errors and improving overall quality of service. 
 
\end{abstract}
\section{Introduction and problem definition}

Customers consuming live TV, video on demand (VoD) or file services are accustomed to a high quality of service which is made possible by content delivery networks (CDN). Modern CDNs usually offer their services via protocols built on top of the HyperText Transfer Protocol (HTTP).The key infrastructure components of CDNs are origin and edge/surrogate servers as well as the communication infrastructure connecting them to their customers. Origin servers contain large volumes of movies, series and other multimedia content and are usually limited in number. Edge servers are more numerous and they are positioned closer to the customers and cache only the most relevant
content due to their limited storage capacity.Various HTTP proxy solutions are used on edge/surrogate servers to cache and serve content to thousands of concurrent users. Software and infrastructure errors occur both on the customer and CDN side. Customer software in set-top boxes, smart televisions and mobile devices might contain bugs or issue erroneous commands.The CDN infrastructure might contain bottlenecks, be under cyber attack or just have partial hardware or software failures. Considering the large number of customers and ever increasing numbers of nodes in modern CDNs, the number of such errors is continuously increasing. 

Log files can provide important insights about the condition of a system or device. These files provide information about whether or not a system is working properly, as well as how actions or services perform. Various types of information are stored in log files on different web servers such as username, timestamp, last page accessed, success rate, user agent, Universal Resource Locator (URL), etc. By analysing these log files, one can better understand user and system behaviour. Since complex system generate large quantities of log information, manual analysis is not practical; thus, automated approaches are warranted \cite{breier2015anomaly}.

The goal of this research is to contribute to the state of the art by analyzing edge server logs collected at a European CDN provider. The error logs analyzed were clustered on the host (i.e., CDN node) level. The ultimate goal of this research was to identify the causes of the diverse errors, provide valuable insights to the operators and developers of the analyzed CDN and thereby contribute towards solving recurring problems and propose potential improvements.

\section{Related works}  

The growing number of Internet users and their increasing demand for the delivery of low-latency content led to the emergence of networks of content delivery networks (CDN) \cite{nygren2010akamai}. A CDN is composed of a variety of points of presence which are essentially proxies containing the most popular content offered on the CDN \cite{brilli2016effect} \cite{rodrigues2013learning}. The core aspect of CDN is the hierarchical design in which top-level, origin servers contain all available content and lower-level nodes contain only the most frequently requested content in a the last mile, i.e., close to customers in a single geographic region \cite{czyrnek2008cdn}.

Similarly to other large-scale enterprise systems, the daily amount of log lines produced by CDNs is measured in the tens or hundreds of millions or maybe even billions. As the manual analysis of such datasets is impractical and often impossible, it is a viable approach to rely on machine learning algorithms which automatically process log lines and discover interesting patterns. One such ML-based approach is error log clustering \cite{grace2011analysis}.

Neha et al \cite{goel2013analyzing}used a web log analyzer tool called Web Log Expert to identify users’ behavior in terms of number of hits, page views visitors and bandwidth. An in-depth analysis of user behaviour of the NASA website was performed in \cite{suneetha2009identifying}. The goal of that research was to obtain information about top errors, websites and potential visitors of the site. The study showed that machine learning techniques such as association, clustering, and classification can be used to identify regular users of a website. The goal of study \cite{johansson2016clustering}  was to determine the best approach for identifying user behavior, whether quantitative or qualitative methods are used. The researchers claimed that clustering users faces two challenges: reasoning and surrounding behavior. They came to the conclusion that combining qualitative and quantitative
methodologies is the best way to understand user behavior. Haifei Xiang in \cite{xiang2020research} used weblog data to cluster user behavior and analyze user access patterns. The study explained how to analyze user behavior from weblog data and apply the K-means clustering algorithm. It considers the disadvantages of K-means to obtain local optimum solutions. Methods such as selecting the initial centers based on data sparsity were proposed, which may effectively minimize algorithm iteration time and increase clustering quality. In \cite{turvcanik2020web} the researchers clustered users into networks according to their browsing behaviour.Their methodology consisted of web log pre-processing and clustering with the K-means algorithm with the ultimate goal to group users into different categories and analyze their behaviour based on the category of the web sites with which they interact.

Log parsing is a technique for converting unstructured content from log messages into a format appropriate for data mining. The aim of the study \cite{turvcanik2020web} was to analyze how log parsing strategies using natural language processing affected log mining performance. The researchers utilized two datasets: the first consisted of log data collected in an aviation system which included over 4,500,000 messages gathered over the period of a year. The second dataset was log data from public benchmarks acquired from a Hadoop distributed file system (HDFS) cluster.

A novel way for presenting textual clustering to automatically find the syntactic structures of log messages logs collected on super-computing systems was proposed in \cite{jain2009extracting}. The researchers managed to utilize their approach to extract meaningful structural and temporal message patterns. The goal of study \cite{sahu2014enhancement} was to investigate user sessions via a frequent pattern mining approach in web logs. 

\section{Data exploration}

The CDN service provider delivered logs in multiple batches,  incrementally improving the data collection and anonymization methods for better fitting to the research goals. This work builds on multiple experimental data sets extracted  from  the base  data set,  which  contains  more  than  2.2  billion  CDN  proxy  log  entries  collected during a 7-day period. This dataset has 29 different features and most of them are categorical data - table  \ref{table:table_features} contains the list of features in the dataset. In the experimentation and prototyping phase of this research work we extracted different sample datasets from the logs. 

Web and web proxy servers send a set of known HTTP status codes when processing a client request results in an error. These errors can be grouped into client (HTTP status codes 4XX) and server errors (HTTP status 5XX). HTTP response status codes show whether or not a particular HTTP request was completed successfully. Responses can be grouped into the following five groups: (1) information response (100-199), (2) successful answers (200-299), (3) redirects (300-399), (4) client errors (400-499), and (5) server errors (500-599).

We considered a log line an error if the HTTP status code logged was equal or greater than 400. We started with the assumption that client side errors if the status code starts with 4 and it will be considered as server side error if the status code starts with 5. 

\begin{table}[H]
\centering
\begin{tabular}{ |m{3cm}|m{10cm}| }
 \hline
 
Feature & Description \\
\hline

statuscode & HTTP response status codes\\
\hline
contenttype & Indicates the media type of the resource\\
\hline
protocol & Http version \\
\hline
contentlength & The size of the resource, in decimal number of bytes\\
\hline
timefirstbyte & time from request processing until the first byte\\
\hline
timetoserv & time needed to process the request\\
\hline
osfamily & Client's device Operating System\\
\hline
sid & id uniq to a streaming session\\
\hline
cachecontrol & Directives for caching mechanisms in both requests and responses\\
\hline
uamajor & User-Agent'version, eg. browser's version\\
\hline
uafamily & User-Agent,usually client’s app \\
\hline
devicefamily & Type of the device \\
\hline
fragment & video or file fragment identifier \\
\hline
path & URL (address of request)\\
\hline
timestamp & Arrival time of the request\\
\hline
contentpackage & VoD asset identifier \\
\hline
coordinates & Long. and lat. of the client based on geoip lookup \\
\hline
livechannel & Live TV channel name \\
\hline
devicemodel & Client's device model \\
\hline
devicebrand & Client's device brand \\
\hline
host & Specifies the domain name of the server (for virtual hosting) \\
\hline
method & Http request method eg. Get,post \\
\hline
manifest & resource which the browser should cache for offline access \\
\hline
assetnumber & VoD asset encoding version \\
\hline
hit & HTTP request was a cache hit or miss \\
\hline
cachename & cache's hostname \\
\hline
popname & cache's location \\
\hline
uid & id unique to a single user \\
\hline

\end{tabular}
\caption{List of log line features}
\label{table:table_features}

\end{table} 

Apart from the HTTP status codes, the log lines contained the following groups of features: (1) user device details (e.g., operating system, device brand, user agent), (2) HTTP request-related information (e.g., was the request a cache hit or not, protocol version, content type and path), (3) CDN node information (host, cache, point of presence - PoP), as well as (4) customer-specific information (geo coordinates, user identifier, session identifier). 

\begin{figure}[H]
  \centering
     \begin{minipage}[b]{1 \textwidth}
    \includegraphics[width=\textwidth,valign=t]{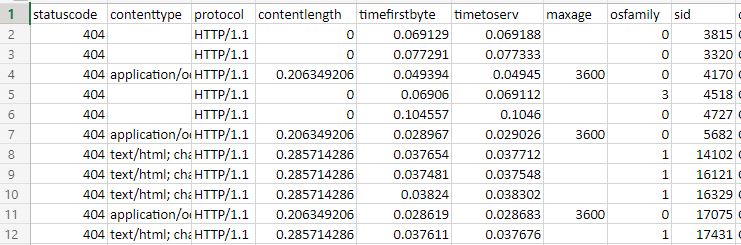}
      \end{minipage}
  \hfill 
  \caption{Sample CDN logs data}
  \label{fig:CDN logs} 
  \end{figure}

\texttt{Figure} \ref{fig:CDN logs} presents a subset the CDN log dataset with a subset of key features.

\section{Feature selection methodology and results}

The data dimensionality problem was solved by utilizing a feature selection technique to reduce the number of features before any experiments. Our goal in this step was to identify the most important features related to the HTTP status codes created by the CDN hosts. For this purpose we used the following four feature selection techniques:

\begin{itemize}
\item  ChiSquare: It works on testing whether a target value depends on input variable or not, all features will be scored and the higher scores indicate higher dependency on the target value \cite{khalid2014survey}.

\item Correlation-based Feature selection: in this technique the linear relationship between variables is analysed and the results indicate the similarity of variables \cite{rachburee2015comparison}. 

\item  ExtraTreeClassifer: This technique classifies the features using a collection of decision trees from a forest. A random sample of features will be extracted from a set of features and in every round the best features will be selected by each decision tree\cite{mochammad2021stable}. 

\item Forward Feature Selection: In this technique an evaluation function will be used and in each iteration one feature will be added to the model then the results will be evaluated. The added feature will be kept in the output set of features if there was an improvement in the results. The process is repeated until the model no longer has any (new) improvements \cite{wah2018feature}.
\end{itemize} 

\begin{figure}[H]
  \centering
     \begin{minipage}[b]{1 \textwidth}
    \includegraphics[width=\textwidth,valign=t]{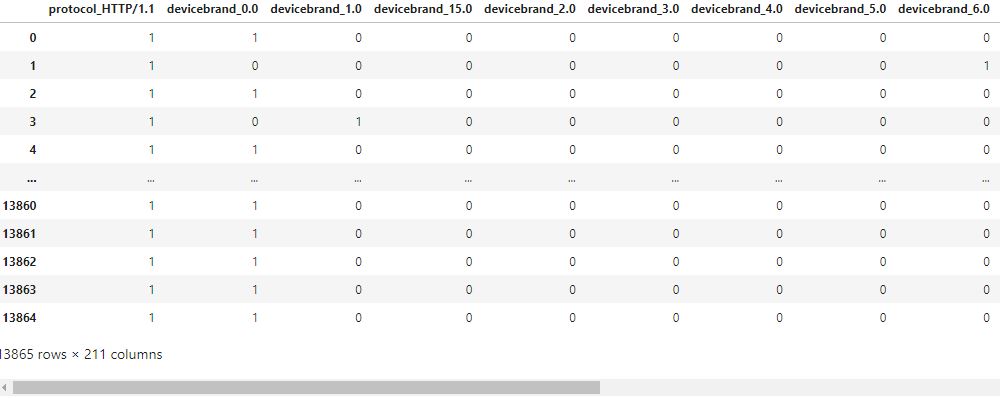}
      \end{minipage}
  \hfill 
  \caption{sample converted categorical data to numerical}
  \label{fig:one hot enchoding} 
  \end{figure}

The goal of the feature selection phase was to reduce the number of attributes in a log record to a minimal representative subset. The one-hot encoding technique was used to transform the categorical features into numerical values - see \texttt{Figure} \ref{fig:one hot enchoding} which visualizes a subset of the resulting dataset.

\subsection{Host-based feature selection}
Each CDN host is responsible for a specific type of data, therefore we might be able to have a better representative of different requests in each host and that can lead us to the source of errors. In order to find the best features for host clustering we applied all techniques listed in the previous section for logs collected on each separate CDN host. Customers can use different services offered by a CDN service provider such as: web content, live TV, video on demand (VOD) and file caching. We chose three hosts with different services, considering the highest number of client error records in each. Our statistical analysis showed that for the live TV service (hosts 1 and 7) and for website service (host 3) had the highest number of error records. Our host-based feature selection showed that the relevant features were the same for all hosts as shown in Table 2. Note that we decided to focus our analysis on exactly those CDN hosts which produced the most error log lines and ignored the other CDN hosts whose logs were present in the dataset.

\begin{table}[H]
\begin{center}
\begin{tabular}{ |p{8cm}|  }
 \hline
 \multicolumn{1}{|c|}{Selected Features for Host 1,3 and 7}\\
 \hline
 statuscode, protocol, contentlength, timefirstbyte timetoserv, osfamily, uamajor, uafamily, devicefamily,path, device brand, method, Live channel\\ 
 \hline
\end{tabular}
 \caption{Selected features for hosts}
\label{table:6}
\end{center}
\end{table}  

\section{Results} 

The goal of the clustering step was to identify a set of error prototypes. Having these prototypes helps focusing on services and parts of CDN infrastructure which produce the most errors.

In this study we used \textit{K-means}, and \textit{K-modes} algorithms \footnote{K-means can be used with categorical data after one-hot encoding.}.   K-modes is one of the most well-known techniques for clustering categorical data \cite{li2020k}. Instead of utilizing the default parameter for  K-means \footnote{https://scikit-learn.org/stable/modules/generated/sklearn.cluster.KMeans.html} and K-modes \footnote{https://pypi.org/project/kmodes/}, we selected the optimal hyper-parameters with grid search. The values of these parameters are shown in tables \ref{table:K-means parameter} and \ref{table:Kmodes parameter} below.

\begin{table}[h!]
\begin{center}
\begin{tabular}{ |p{5cm}|p{5cm}| }
 \hline
 
\textbf{Parameters} & \textbf{Values}  \\\hline
 
\texttt{n\_clusters} & 6 \\\hline
\texttt{init} & k-means++\\\hline
\texttt{max\_iter} & 300 \\\hline
\texttt{n\_init} & 10\\\hline
\texttt{random\_state} & 0\\\hline  
\end{tabular}
 \caption{Parameters used for K-Means Clustering}
\label{table:K-means parameter}
\end{center}
\end{table}

\begin{table}[h!]
\begin{center}
\begin{tabular}{ |p{5cm}|p{5cm}| }
 \hline
 
\textbf{Parameters} & \textbf{Values}  \\\hline
 
\texttt{n\_clusters} & 8 \\\hline
\texttt{init} & Huang\\\hline
\texttt{n\_init} & 5\\\hline
\texttt{verbose} & 1\\\hline  
\end{tabular}
 \caption{Parameters used for Kmodes Clustering}
\label{table:Kmodes parameter}
\end{center}
\end{table}

\subsection{Host-specific error clustering}
Hosts correspond to distinct services offered by the CDN network, e.g., video on demand, live TV or file services. We filtered the dataset and extracted all log lines with \texttt{statuscode}s $\geq 400$. 

In the dataset we identified log records from 57 hosts, but only 20 were actually valid hosts, the rest were created for different processes such as testing. After analyzing the valid hosts we found out that most of them only have a small number of records which can be easily analysed without clustering. Accordingly we only selected those hosts that had more than $1000$ data points. As a result we ended up having three valid hosts $1$ (live TV), $5$ (website) and $7$ (live TV). At this point we created three different sample data sets which only include the records from each specific host. Next we analysed them with different clustering algorithms.

For these experiments we used the features selected from Section 4.1, then we applied the K-Means and K-modes clustering algorithms.  

\subsection{Host 1 - Live TV}

In our first experiment we used K-means with one hot encoding of categorical features \footnote{https://towardsdatascience.com/what-is-one-hot-encoding-and-how-to-use-pandas-get-dummies-function-922eb9bd4970}. The total number of data points for Host 1 was $13886$ which were grouped into $6$ different clusters. Most of the errors in this host were server errors, with a smaller number of client errors present.

\begin{figure}[H]
  \centering
  \begin{minipage}[b]{0.47\textwidth}
    \includegraphics[width=\textwidth,valign=t]{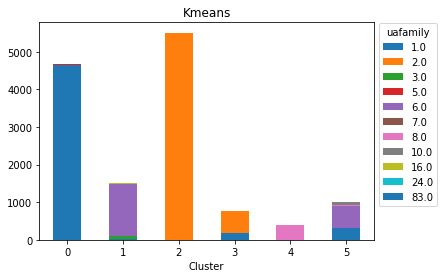} 
  \end{minipage}
  \hfill
  \begin{minipage}[b]{0.49\textwidth}
    \includegraphics[width=\textwidth,valign=t]{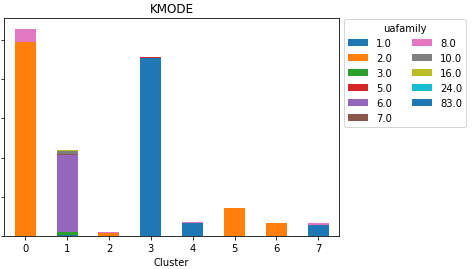} 
  \end{minipage}
      \caption{Host 1 -Cluster using K-means and K-mode}
    \label{fig:host1}
  \end{figure}

Table \ref{table:host1_one_hot_kmeans} and figure \ref{fig:host1} show the error log clusters for host 1.

\begin{table}[H]
\begin{center} 
\begin{tabular}{ |p{1 cm}|p{1.7cm}|p{3cm}|p{7cm}|  }
 \hline
 \multicolumn{4}{|c|}{K-means One hot encoding for host 1} \\
 \hline
 cluster & datapoints &  status code  & Relationship with other attributes  \\\hline
 
0&	4676&	412 (5 records)    500 (37 records)    502 (4634 records) &	Protocol =HTTP 1.1, device brand =0,2,3,4,5,8 , Method =GET, device family=1,4,6,10,13,24,38,49,51,58,66,36,650, uafamily=1,5, Live channel =many , osfamily =1 \\\hline
1&	1523 &	403 (12 records) 502 (1510 records) 503 (1 records)  &	Protocol =Http 1.1 , device brand =nan, method =GET , device family =0 ,nan, uafamily=1,3,6,7,16,24,nan , live channel =many, os family=0,2,3,4,nan\\\hline
2& 5508 &412 (15 records) 500 (82 records) 502 (5411 records) & Protocol =Http 1.1, device brand =0, device family=1, ua family =2, live channel=many, os family= 1 \\\hline
3&	753 &	500 (17 records)  502 (736) &	Protocol =Http1.1, device brand= 2,5,6, method =GET, device family =many, ua family= 1,2, live channel =many, os family = 1 \\\hline
4&	389  &	500 (1 records) 502 (388 records) &	Protocol =Http 1.1, device brand = 4, method= GET, device family =8, ua family = 8, live channel = many, os family = 6\\\hline
5&	1016 &	403 (9 records) 416 (26 records) 502 (981 records) &	Protocol =Http 1.1, device brand =1, method=GET, device family =2,5,7,9,11, ua family= 1,6,8,10, live channel= many, os family=3,5,7 \\
 \hline
\end{tabular}
 \caption{Host 1 clustering using K-Means}
\label{table:host1_one_hot_kmeans}
\end{center}
\end{table}

We clustered the error logs collected for host 1 with the K-Mode algorithm and with one hot encoding. The results were very similar to the previous experiment, one difference was that here we had $8$ clusters. As can be seen in \texttt{Figure} \ref{fig:host1} this host had a significant number of server errors $502$ which shows that it most probably suffered from a traffic overload. Further analyses is warranted to investigate the temporal aspect of that overload. 
 
 \begin{figure}[H]
  \centering
     \begin{minipage}[b]{0.48\textwidth}
    \includegraphics[width=\textwidth,valign=t]{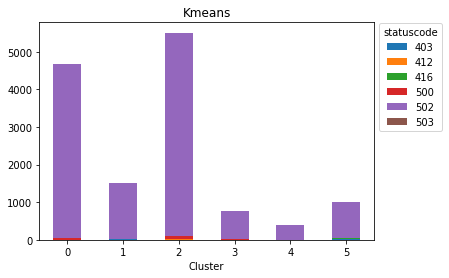}
      \end{minipage}
  \hfill
  \begin{minipage}[b]{0.49\textwidth}
    \includegraphics[width=\textwidth,valign=t]{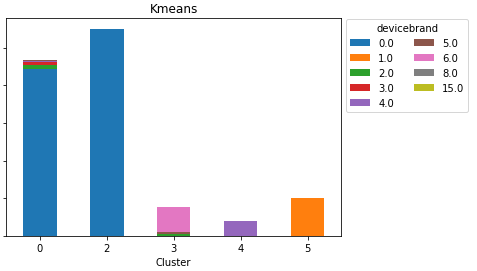}
  \end{minipage}
  
  \end{figure}
  
  \begin{figure}[H]
  \centering
  \begin{minipage}[b]{0.45\textwidth}
    \includegraphics[width=\textwidth,valign=t]{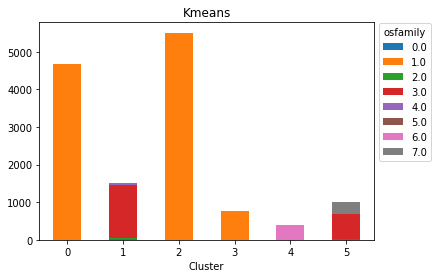}
  \end{minipage} 
  \hfill
  \begin{minipage}[b]{0.50\textwidth}
    \includegraphics[width=\textwidth,valign=t]{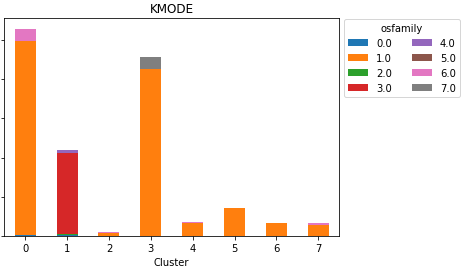} 
     \end{minipage}
     \vspace{-20pt}
   \end{figure}
    \begin{figure}[H]
  \centering
  \begin{minipage}[b]{0.47\textwidth}
    \includegraphics[width=\textwidth,valign=t]{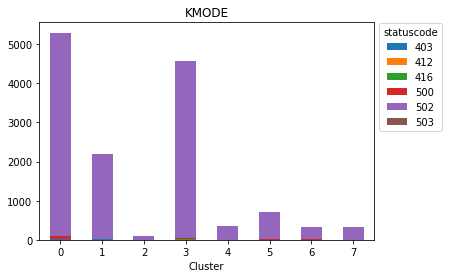} 
     \end{minipage}
  \hfill
  \begin{minipage}[b]{0.49\textwidth}
    \includegraphics[width=\textwidth,valign=t]{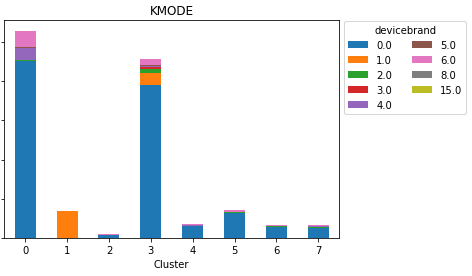} 
      \end{minipage}
    \end{figure}

\subsection{Host 3 - Web content}
There were $1235$ errors logged on Host 3, $1197$ forbidden requests (HTTP \texttt{statuscode} $403$) and $40$ method not allowed errors (HTTP \texttt{statuscode} $405$). A total of $7$ clusters were identified with k-means and $9$ clusters with K-modes.

\begin{figure}[H]
  \centering
  \begin{minipage}[b]{0.43\textwidth}
    \includegraphics[width=\textwidth,valign=t]{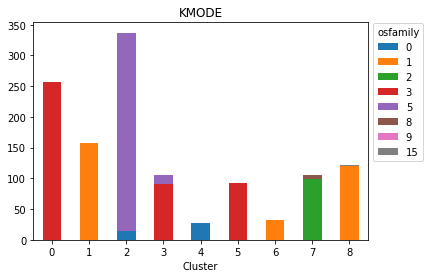} 
  \end{minipage}
  \hfill
  \begin{minipage}[b]{0.52\textwidth}
    \includegraphics[width=\textwidth,valign=t]{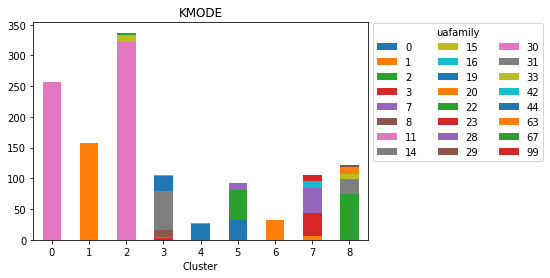} 
    
  \end{minipage}
  \caption{Host 3 -Cluster using K-means and K-mode}
  \label{fig:host3}
  \end{figure}

These errors corresponded to customers using different \texttt{devicefamily, devicebrand, osfamily} and \texttt{uafamily} devices. The feature distribution in each cluster is shown in \texttt{Figure} \ref{fig:host3}. They were most probably caused by customer device bugs or malicious activity. Note that clusters 3, 4 and 7 are not shown in the bar charts in \texttt{Figure} \ref{fig:host3} due to a small number of the data points.  

\vspace{40pt}
\setlength\intextsep{0pt}
\begin{figure}[H]
  \centering 
  \begin{minipage}[b]{0.46\textwidth}
    \includegraphics[width=\textwidth,valign=t]{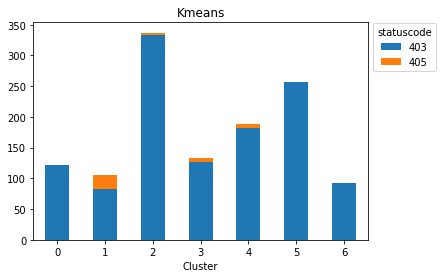}
  \end{minipage}
  \hfill 
  \begin{minipage}[b]{0.50\textwidth}
    \includegraphics[width=\textwidth,valign=t]{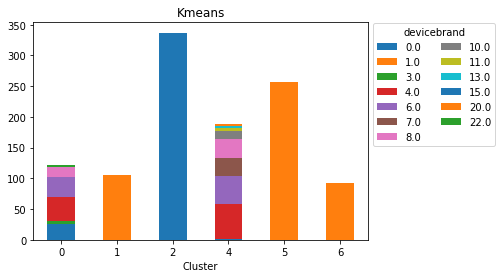}
  \end{minipage}
  \end{figure}
  
  \begin{figure}[H]
  \centering 
  \begin{minipage}[b]{0.43 \textwidth}
  \includegraphics[width=\textwidth,valign=t]{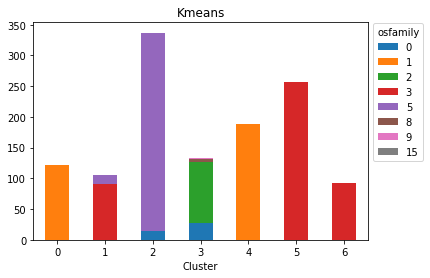}
  \end{minipage}
  \hfill 
  \begin{minipage}[b]{0.50\textwidth}    \includegraphics[width=\textwidth,valign=t]{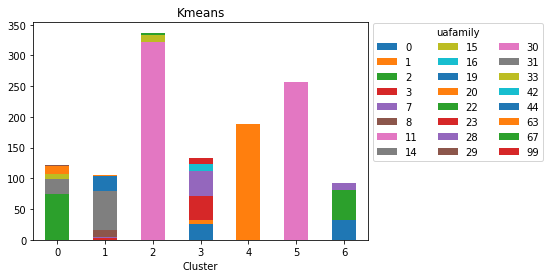} 
    
  \end{minipage}
  
  \end{figure}
   \begin{figure}[H]
  \centering
  \begin{minipage}[b]{0.45\textwidth}
    \includegraphics[width=\textwidth,valign=t]{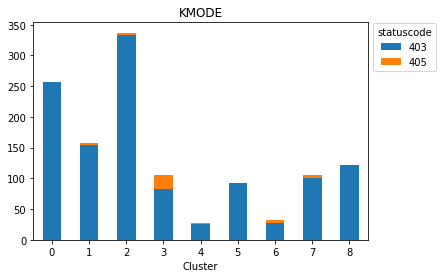} 
  \end{minipage}
  \hfill
  \begin{minipage}[b]{0.50\textwidth}
    \includegraphics[width=\textwidth,valign=t]{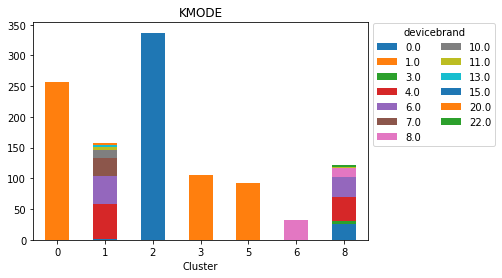} 
    
  \end{minipage}
  
  \end{figure}

\subsection{Host7 - Live TV}
Host 7 logged $31492$ error records. $20272$ bad requests (HTTP status code $400$), $11078$ precondition failed (HTTP $412$), $129$ service unavailable (HTTP $503$) and $12$ bad requests (HTTP $403$). Our clustering approach grouped these records into 7 clusters. 

\begin{figure}[H]
  \centering
  \begin{minipage}[b]{0.34\textwidth}
    \includegraphics[width=\textwidth,valign=t]{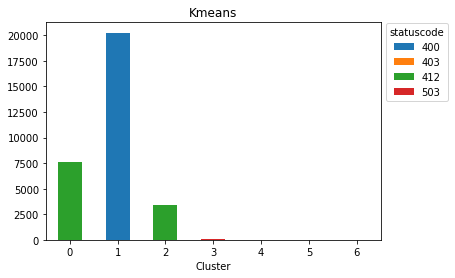}
  \end{minipage}
  \hfill
  \begin{minipage}[b]{0.3\textwidth}
    \includegraphics[width=\textwidth,valign=t]{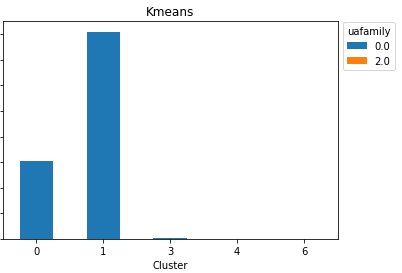}
  \end{minipage}
  \hfill
  \begin{minipage}[b]{0.3\textwidth}
    \includegraphics[width=\textwidth,valign=t]{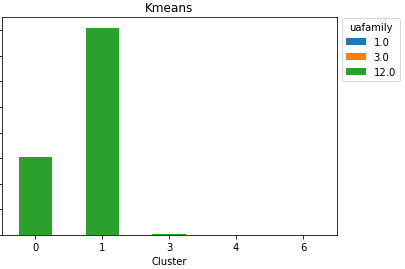}
  \end{minipage}
 
  \hfill 
  \begin{minipage}[b]{0.35\textwidth}
    \includegraphics[width=\textwidth,valign=t]{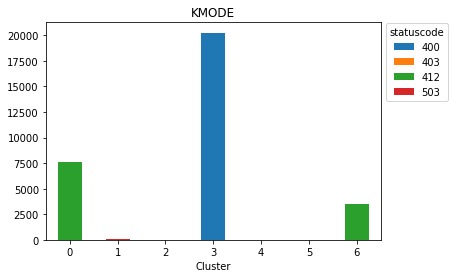} 
    
 \end{minipage} 
  \hfill
  \begin{minipage}[b]{0.3\textwidth}
    \includegraphics[width=\textwidth,valign=t]{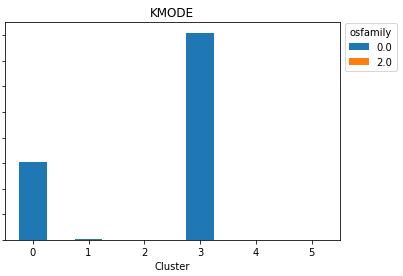} 
    
  \end{minipage}  
  \hfill 
   \begin{minipage}[b]{0.3\textwidth}
    \includegraphics[width=\textwidth,valign=t]{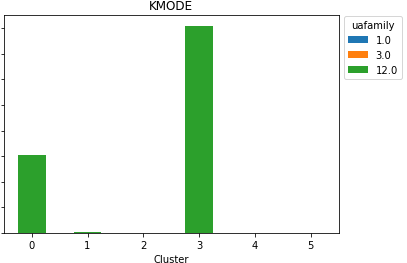} 
    
  \end{minipage}  
  \hfill
  
  \caption{Host 7 -Cluster using K-means and K-mode}
   \label{fig:host7}
\end{figure}

In this host we noticed that the number of errors was significantly larger compared to Host 1 and 3 and all records corresponded to the same path with the same \texttt{device family, uafamily} and \texttt{osfamily}. After investigating we found out that these are actually a result of crawling. \texttt{Figure} \ref{fig:host7} is an illustration of different features in each clusters. 

\subsection{Discussion}

Error log clustering in content delivery networks is a less-frequently analysed topic, although it might allow CDN operators to pinpoint frequent error clusters and allow them to mitigate their causes. In that context, the log analysis approach prestented in this paper provides valuable novel information to CDN operators. We found that the CDN hosts analyzed in our experiments tended to produce different sets of errors. Host 3, a web server, only logged client errors $403$ and $405$, which might result from bugs in customer devices/browsers. On live TV host 1 the vast majority of errors were produced by the CDN nodes (HTTP status code $\geq 500$) which might point to a software issues or an overload on that specific host.

We were not completely satisfied with the results of our clustering approach, as ideally we would like to see clusters that clearly separate the different kinds of errors closest to the error prototypes (i.e., cluster centroids). We found that some of the identified clusters contained multiple HTTP error codes. This does not necessary mean that the clustering is bad. When we examined the sessions at the different hosts, we discovered that even the very same sessions can produce multiple error codes. Another, simpler explanation might be, that same type of resources just cause different errors on different resources, as server errors when something goes wrong within the CDN infrastructure, or client errors when someone simply provides a non existing URL. These explanations however naturally require further investigations.

\section{Conclusion and Future work }

The goal of this research was to apply different clustering techniques for mining value-added information from real-life CDN logs. We worked with a dataset consisting from 2.2 billion log lines collected over a 1-week period on multiple CDN hosts offering video on demand, web content, file and live TV services. 

In the data exploration phase we filtered out the log lines corresponding to HTTP error codes ($\geq 400$). We performed feature selection with different techniques and thereby reduced the total number of features from 29 to 10. Additionally, we decided to separately analyze the error logs corresponding to hosts offering video on demand, live TV and web content. We experimented with the K-means and K-mode algorithms and identified meaningful and explainable error clusters, which can be utilized by CDN operators to obtain increased situational awareness and optimize the workings of their systems. We also introduced a novel, visual representation method (figures \ref{fig:host1}, \ref{fig:host3} and \ref{fig:host7}) which visually conveys information about clusters of observations described by multiple categorical features.

We analyzed all error logs collected over a period of a single week on each of the CDN hosts, which means that our analysis was time-agnostic. In future we plan to incorporate the temporal aspect into our analysis. Additionally, we also plan to perform user session-level error analysis and thereby better understand why some communication sessions end in (HTTP) errors.

%
%

%
%
%
%
\bibliographystyle{plain}
\bibliography{bibliography.bib}

\end{document}